\documentclass[10pt,letterpaper]{article}
\usepackage[utf8]{inputenc}
\usepackage{amsmath}
\usepackage{amsfonts}
\usepackage{amssymb}
\usepackage{multirow}
\usepackage{bm}
\usepackage{xcolor}
\usepackage[english]{babel}
\usepackage{graphicx}
\usepackage{hyperref,cite}
\hypersetup{
    colorlinks=true,
    linkcolor=blue,
    filecolor=blue,  
    citecolor=blue,
    urlcolor=blue,
    }
\usepackage{lipsum}
\usepackage{multicol}

\title{Lagrangian and Multisymplectic Descriptions of Classical Fields: A Connection in the Momentum Representation}

\author{J. F. Pérez-Barragán\footnote{Contact email address: jfcoperezb@estudiantes.fisica.unam.mx}\\
\normalsize{Instituto de Física, Universidad Nacional Autónoma de México, 04510 Mexico City, Mexico}}

\date{}

\begin{document}

\maketitle

\abstract{\noindent The multisymplectic Hamiltonian formalism is a generalization of the Hamiltonian formalism that manifestly preserves covariance in the description of fields and that has been proposed as a possible framework for developing a Lorentz-covariant, canonical quantization scheme. However, the possibility of defining multiple Poisson brackets within this formalism has significantly limited its practical use in field theory. In this paper, we establish a connection between the Lagrangian and multisymplectic descriptions of classical fields interacting with point particles in the field's momentum representation by deriving the image of the de Donder–Weyl function for a general tensor representation of the Lorentz group. The calculation is carried out explicitly for the complex scalar field, the electromagnetic field, and the classical Dirac field. On this basis, we propose a Lorentz-covariant Poisson bracket with respect to the canonical variables of the fields, thereby opening the possibility of studying the complete system within a consistent relativistic and canonical framework.

\vspace{0.25cm}
}

\section{Introduction}

The treatment of fields as mechanical systems with an infinite number of degrees of freedom allows formulating field theories by pure analogy to the mechanical description of particles. Hence, there are two methods for constructing a field theory: the Hamiltonian and the Lagrangian formalisms. The usual formulation of the Hamiltonian (or canonical) formalism is characterized by the primacy of time as the fundamental independent variable and thus generates field theories that are devoid of manifest covariance, so this property must be verified at every stage of construction, typically through laborious calculations. In the Lagrangian formalism, on the other hand, all space-time coordinates intervene symmetrically; therefore, this formalism implies field theories in full agreement with the principle of relativity. Then, since the main objective of all field theories, both classical and quantum, is to locally describe the interactions between particles, modern field theories have as a basic requirement to agree with the special theory of relativity and hence are usually developed on the basis of the Lagrangian formalism. Nevertheless, the most recurrent approach to formulate quantum field theories is through the so-called canonical quantization process, which is entirely based on a Hamiltonian treatment. Naturally, this incompatibility is only apparent, since quantum field theories derived using canonical quantization correctly describe relativistic phenomena; however, it raises the question on whether the Hamiltonian formalism can be generalized in such a way as to allow for a manifestly Lorentz-covariant, canonical quantization process.

The multisymplectic (or multiphase) Hamiltonian formalism (MHF) emerges in that sense as an alternative that extends the concepts and methods of the Hamiltonian formulation of classical mechanics to the realm of field theories in a Lorentz-covariant manner by introducing additional generalized momenta and defining a generalized, Lorentz-invariant Hamiltonian density: the de Donder–Weyl function. Although the basic proposal behind this generalization can be tracked back to the work of de Donder \cite{Donder1935} and Weyl \cite{HW1935} on the geometrical formulation of the calculus of variations in the 1930s, the potential of the MHF was not recognized until the 1970s, when a compact, but active group of researchers found it useful in the context of the general theory of relativity (see, for example, \cite{KG1972,JK1973,JK1976,WS1976,JK1979} and the references cited therein). However, the general interest in this formalism began in 1986 with the exploratory work of Marsden et al. on the possibility of constructing Lorentz-covariant Poisson brackets in different field theories \cite{JM1986} and the subsequent development of multisymplectic descriptions in hydrodynamics and the theory of elasticity \cite{JM1999,JM2001}. In recent years, the interest in the MHF as a possible way to quantize field theories in a Lorentz-covariant manner has gained renewed momentum, leading, among other results (see \cite{IK1998,IK1999,MN1998,IK2001b,IK2004,FH2015,IK2015,VG2006}, for example), to a first series of proposals for constructing a quantum theory of gravity \cite{IK2000,IK2001a} and to a relativistic generalization of Bohmian quantum mechanics \cite{HN2005,HN2006}.

Having said that, a question that immediately arises is: why is it not customary to use the MHF to construct field theories? The answer lies in the fact that the development of this formalism beyond its basic definitions is not straightforward and therefore still has major formal lacunae. In particular, the MHF allows for multiple alternatives for defining essential objects, such as Poisson brackets, which leads to ambiguities in establishing the algebraic structure of the extended phase space \cite{IK1997,MM1998,IK1998a,MM2001,JS2008}. Additionally, the physical interpretation of the different terms that appear in the multisymplectic descriptions of systems is yet a matter of debate \cite{LR2020}. Furthermore, the association of each canonical position with more than one momentum immediately results in the loss of the canonical duality between positions and momenta, which is crucial for establishing general quantization schemes \cite{MF2002,MF2005,MF2015}. Consequently, the discussions on the MHF have remained largely confined to the realm of mathematics.

Still, recent investigations by Cetto and others, exploring the connection between the matrix formulation of quantum mechanics and the response of a particle to the zero-point radiation field, i.e., the non-thermal, stochastic, electromagnetic background persisting even in the absence of matter, motivate to revisit the problem of constructing a canonical quantization process through the MHF, since that connection is established by expressing the particle's canonical variables in terms of the normal variables of the field's modes of oscillations by means of Poisson brackets. So, in the stationary regime of the particle-field interaction, when the field effectively controls the particle’s dynamics, the kinematical transformation that changes the meaning of the physical quantities describing the particle becomes manifest \cite{Cetto2021,Cetto2022a,Cetto2022b,Cetto2023}. Hence, a relativistic generalization of Cetto’s approach to quantization to field theories would not only provide a physical rationale for replacing phase-space functions with operators in the description of fields, but also shed light on how to formulate a Lorentz-covariant, canonical quantization procedure. 

The aim of the present article is thus to establish a connection between the Lagrangian and multisymplectic descriptions of classical fields after their decomposition into normal modes of oscillation, so as to enable a unified, Lorentz-covariant, canonical treatment of fields and particles, and, thereby, provide a basis for extending the results of Cetto in the aforementioned manner. Therefore, after a brief overview of the foundations of the MHF, we immediately develop within this formalism a linear field theory for a general tensor field interacting with point particles. We then proceed to transition the resulting description to the field's momentum representation by two different methods and establish the correspondence between the two formalisms. This procedure is also carried out explicitly for the complex scalar field and the electromagnetic field as illustrative examples. Subsequently, we explore the scope of the MHF by analyzing a system composed of spinor fields. The paper ends with a general proposal for developing a Lorentz-covariant, canonical description of a system of point particles interacting with an electromagnetic radiation field by means of a Poisson bracket jointly containing the canonical variables of the field and the mechanical subsystem. It is noteworthy that when dealing exclusively with local field theories, that is, field theories with first-order Lagrangian densities, it is not necessary to use the formalism of differential geometry, so we will be able to maintain the mathematical language throughout the paper within the framework of tensor calculus.

\section{The Multisymplectic Hamiltonian Formalism}

Let us consider a four-dimensional Minkowski space-time, with associated metric tensor $\eta_{\mu\nu}$ and signature $(+,-,-,-)$, and a field theory defined by the action
\begin{equation}
    S=\int{d}^{4}x\hspace{0.5mm}\mathcal{L}(x_{\mu},\phi_{a},\partial_{\mu}\phi_{a}),
\end{equation}
where $\mathcal{L}(x_{\mu},\phi_{a},\partial_{\mu}\phi_{a})$ is a first-order Lagrangian density depending on a collection of classical fields $\phi_{a}$, with $a=1,...,N$. Then, a manifestly Lorentz-covariant, canonical description of the system can be achieved by associating each field $\phi_{a}$ with the generalized (multisymplectic) momentum
\begin{equation}
    \theta_{a\mu}=\frac{\partial\mathcal{L}}{\partial(\partial^{\mu}\phi_{a})},\label{MultiMomentumDef}
\end{equation}
since this association allows rewriting the action as
\begin{equation}
    \int{d}^{4}x\hspace{0.5mm}\left(\sum_{a=1}^{N}\theta_{a\mu}\partial^{\mu}\phi_{a}-\mathcal{H}(x_{\mu},\phi_{a},\theta_{a\mu})\right),\label{LegendreTrans}
\end{equation}
with
\begin{equation}
    \mathcal{H}(x_{\mu},\phi_{a},\theta_{a\mu})=\sum_{a=1}^{N}\theta_{a\mu}\partial^{\mu}\phi_{a}-\mathcal{L}(x_{\mu},\phi_{a},\partial_{\mu}\phi_{a})\label{LegendreTrans2}
\end{equation}
a Lorentz-invariant function of space-time coordinates, fields, and multisymplectic momenta that is occasionally referred to as de Donder–Weyl function \cite{Donder1935,HW1935}. It is important to point out that in order to establish the last equation, we have assumed that the Lagrangian density is regular, i.e., that the determinant of the corresponding Hessian matrix, $\partial^{2}\mathcal{L}/\partial(\partial_{\mu}\phi_{a})\partial(\partial_{\nu}\phi_{a})$, is non-vanishing for each field. As in the Hamiltonian formulation of classical mechanics, this requirement ensures the independence between field derivatives and multisymplectic momenta. Examples of field theories with singular Lagrangian densities are examined within the MHF framework in Secs. \ref{CampEM} and \ref{Dirac}.

On this basis, if the variations of fields and multisymplectic momenta are treated independently, it follows that the action in Eq. (\ref{LegendreTrans}) leads to the variation
\begin{equation}
    \delta{S}=\int{d}^4x\sum_{a=1}^{N}\left(\left[\partial_{\mu}\phi_{a}-\frac{\partial\mathcal{H}}{\partial\theta_{a}^{\mu}}\right]\delta\theta_{a}^{\mu}-\left[\partial_{\mu}\theta_{a}^{\mu}+\frac{\partial\mathcal{H}}{\partial\phi_{a}}\right]\delta\phi_{a}+\partial_{\mu}[\theta_{a}^{\mu}\delta\phi_{a}]\right),
\end{equation}
so, by adopting the requirement of nullity of variations of fields at infinity, one immediately obtains that the action is stationary only if equations
\begin{equation}
    \partial_{\mu}\phi_{a}=\frac{\partial\mathcal{H}}{\partial\theta_{a}^{\mu}},\quad\partial_{\mu}\theta_{a}^{\mu}=-\frac{\partial\mathcal{H}}{\partial\phi_{a}}\label{EqHamiltonPi}
\end{equation}
are satisfied. These equations are known as the covariant Hamilton's equations and fully determine the dynamics of the system in accordance with Hamilton’s principle of least action. Additionally, it should be emphasized that the covariance of the multisymplectic description is ensured from the outset, since the de Donder–Weyl function inherits the Lorentz-invariant condition from the Lagrangian density via the Legendre transformation in Eq. (\ref{LegendreTrans2}) \cite{JS2008,MF2002,MF2005,MF2015}.

\section{Multisymplectic Description of a General Tensor Field}\label{DevelopGral}

An arbitrary system composed of interacting classical fields and point particles is described by an action of the form
\begin{equation}
    S=\int{d}^{4}x\hspace{0.5mm}(\mathcal{L}_{\mathrm{part}}+\mathcal{L}_{\mathrm{field}}+\mathcal{L}_{\mathrm{int}}),\label{action}
\end{equation}
with $\mathcal{L}_{\mathrm{part}}$, $\mathcal{L}_{\mathrm{field}}$, and $\mathcal{L}_{\mathrm{int}}$ the Lagrangian densities associated with the free motion of the point particles, the free propagation of the fields, and the interaction between particles and fields, respectively. In the present work, the specific choice of a Lagrangian density for the free point particles will not be of particular importance, since we shall focus on the dynamics of the fields. Furthermore, neither  $\mathcal{L}_{\mathrm{part}}$ nor $\mathcal{L}_{\mathrm{int}}$ depend on the derivatives of the fields when considering a minimum coupling, so the derivation of the conjugate momenta and the Hamiltonian density of the point particles does not require the use of the MHF.

For a complex, massive, $\ell$-th-order tensor field $T_{\nu_{1}\cdots\nu_{\ell}}$, it is straightforward to establish that the Lagrangian density associated with its free propagation is given by
\begin{equation}
    \mathcal{L}_{\mathrm{field}}=a^{2}\partial_{\mu}T^{\ast}_{\nu_{1}\cdots\nu_{\ell}}\partial^{\mu}T^{\nu_{1}\cdots\nu_{\ell}}-b^{2}T^{\ast}_{\nu_{1}\cdots\nu_{\ell}}T^{\nu_{1}\cdots\nu_{\ell}},\label{LTensorField}
\end{equation}
where $a^2$ and $b^2$ are appropriate proportionality constants satisfying that $b^{2}/a^{2}=\kappa^{2}$, with $\kappa$ a parameter having dimensions of inverse length. Moreover, based on Bergmann's proposal for the scalar field \cite{OB1956} and the form of the interaction between charged point particles and the electromagnetic field, we propose for the tensor interaction Lagrangian density that
\begin{equation}
    \mathcal{L}_{\mathrm{int}}=-T_{\nu_{1}\cdots\nu_{\ell}}\sum_{j=1}^{n}g_{j}\int{d}\tau_{j}\hspace{0.5mm}\dot{u}^{\nu_{1}}_{j}\cdots\dot{u}^{\nu_{\ell}}_{j}\delta^{(4)}(x_{\alpha}-u_{j\alpha})-\mathrm{c.c.},\label{InterDensityTensor}
\end{equation}
where we have introduced $g_{j}$, $\tau_{j}$, and $u_{j\mu}$ as the coupling factor with the field, the proper time, and the position of the $j$-th point particle, respectively.

Hence, we obtain by substituting the Lagrangian density in Eq. (\ref{LTensorField}) into Eq. (\ref{MultiMomentumDef}) that the multisymplectic momenta of the system are given by
\begin{equation}
    \theta_{\mu\nu_{1}\cdots\nu_{\ell}}=a^{2}\partial_{\mu}T^{\ast}_{\nu_{1}\cdots\nu_{\ell}}\label{DefMomMulti}
\end{equation}
and its complex conjugate, and that the de Donder–Weyl function thereby is
\begin{equation}
    \mathcal{H}=\frac{1}{a^{2}}\theta^{\ast}_{\mu\nu_{1}\cdots\nu_{\ell}}\theta^{\mu\nu_{1}\cdots\nu_{\ell}}+b^{2}T^{\ast}_{\nu_{1}\cdots\nu_{\ell}}T^{\nu_{1}\cdots\nu_{\ell}}-\mathcal{L}_{\mathrm{int}}.\label{DWFunctionTensor}
\end{equation}
We thus deduce directly from $\mathcal{H}$ that the covariant Hamilton’s equations for the tensor field are
\begin{equation}
    \partial_{\mu}T_{\nu_{1}\cdots\nu_{\ell}}=\frac{1}{a^{2}}\theta^{\ast}_{\mu\nu_{1}\cdots\nu_{\ell}},\label{EcDW1Tensor}
\end{equation}
\begin{equation}
    \partial^{\mu}\theta_{\mu\nu_{1}\cdots\nu_{\ell}}=-b^{2}T^{\ast}_{\nu_{1}\cdots\nu_{\ell}}-\sum_{j=1}^{n}g_{j}\int{d}\tau_{j}\hspace{0.5mm}\dot{u}_{j\nu_{1}}\cdots\dot{u}_{j\nu_{\ell}}\delta^{(4)}(x_{\alpha}-u_{j\alpha}),\label{EcDW2Tensor}
\end{equation}
and their complex conjugates. As in the usual Hamiltonian formalism, Eq. (\ref{EcDW1Tensor}) and its complex conjugate give back the definitions of the multisymplectic momenta. On the other hand, by deriving Eq. (\ref{EcDW1Tensor}) and substituting the complex conjugate of Eq. (\ref{EcDW2Tensor}) into the result obtained, we have that the field equation for the tensor field is given by an inhomogeneous Klein–Gordon equation for each component of the field, as it would
be obtained from the Euler–Lagrange equation of the Lagrangian formalism for the action in Eq. (\ref{action}).

\section{Transition to the Field's Momentum Representation}\label{DescripMomSpace}

The deduction of the correct field equation for the tensor field shows the equivalence between the Lagrangian and the multisymplectic descriptions of systems at the level of the dynamics. Let us now go beyond the basic framework of the MHF by transitioning the above description to the momentum space of the field. Hence, we firstly note that from the method of Green's functions, the retarded solution of the field equation can be expressed as
\begin{equation}
    T_{\nu_{1}\cdots\nu_{\ell}}=\frac{1}{8\pi^{3}}\int{d}^{4}k\hspace{0.5mm}\Theta(k_{0})\delta({k}_{\mu}k^{\mu}-\kappa^{2})\left[\Tilde{T}_{+\hspace{0.5mm}\nu_{1}\cdots\nu_{\ell}}+\Tilde{T}_{-\hspace{0.5mm}\nu_{1}\cdots\nu_{\ell}}\right],
\end{equation}
where the Dirac $\delta$-function $\delta({k}_{\mu}k^{\mu}-\kappa^{2})$ and the Heaviside step function $\Theta(k_{0})$ ensure the fulfillment of the dispersion relation for the tensor field, and $\Tilde{T}_{\pm\hspace{0.5mm}\nu_{1}\cdots\nu_{\ell}}$ are two independent, complex, $\ell$-th-order tensor amplitudes that are given by
\begin{equation}
    \Tilde{T}_{\pm\hspace{0.5mm}\nu_{1}\cdots\nu_{\ell}}=\left(\Tilde{T}_{\pm\hspace{0.5mm}\nu_{1}\cdots\nu_{\ell}}^{\hspace{0.5mm}\mathrm{free}}\mp\frac{i}{a^{2}}\sum_{j=1}^{n}g_{j}\int{d}\tau_{j}\hspace{0.5mm}\dot{u}_{j\nu_{1}}\cdots\dot{u}_{j\nu_{\ell}}\Theta(x_{0}-u_{j0})e^{\pm{i}k_{\mu}u^{\mu}_{j}}\right)e^{\mp{i}k_{\mu}x^{\mu}},\label{AmplTensor}
\end{equation}
with $\Tilde{T}_{\pm\hspace{0.5mm}\nu_{1}\cdots\nu_{\ell}}^{\hspace{0.5mm}\mathrm{free}}$ tensor amplitudes depending solely on $k_{\mu}$. Eq. (\ref{AmplTensor}) reveals that the presence of point particles results for the field in a superposition of plane waves with constant amplitudes that are associated with the field's free propagation and of waves with variable amplitudes whose origin is the interaction between field and particles.

Subsequently, from the explicit forms of $\Tilde{T}_{\pm\hspace{0.5mm}\nu_{1}\cdots\nu_{\ell}}$, we obtain that these amplitudes satisfy the first-order differential equation
\begin{equation}
    (k_{\alpha}\partial^{\alpha}\pm{i}{k}_{\alpha}k^{\alpha})\Tilde{T}_{\pm\hspace{0.5mm}\nu_{1}\cdots\nu_{\ell}}=\mp\frac{ik_{0}}{a^{2}}\sum_{j=1}^{n}g_{j}\int{d}\tau_{j}\hspace{0.5mm}\dot{u}_{j\nu_{1}}\cdots\dot{u}_{j\nu_{\ell}}\delta(x_{0}-u_{j0})e^{\mp{i}k_{\mu}(x^{\mu}-u^{\mu}_{j})}\label{EcAmplTensor}
\end{equation}
and that this equation can be decompose into the pair of first-order differential equations with real coefficients
\begin{multline}
    \partial_{\mu}q_{\pm\hspace{0.5mm}\nu_{1}\cdots\nu_{\ell}}=\pi_{\pm\hspace{0.5mm}\mu\nu_{1}\cdots\nu_{\ell}}+\frac{2\epsilon{k}_{0}}{a^{2}}\frac{k_{\mu}}{k_{\alpha}k^{\alpha}}\sum_{j=1}^{n}g_{j}\int{d}\tau_{j}\hspace{0.5mm}\dot{u}_{j\nu_{1}}\cdots\dot{u}_{j\nu_{\ell}}\delta(x_{0}-u_{j0})\\
    \times\left[\mathrm{Re}(z)\cos(k_{\alpha}(x^{\alpha}-u^{\alpha}_{j}))+\mathrm{Im}(z)\sin(k_{\alpha}(x^{\alpha}-u^{\alpha}_{j}))\right],\label{EqReal1}
\end{multline}
\begin{multline}
    \partial^{\mu}\pi_{\pm\hspace{0.5mm}\mu\nu_{1}\cdots\nu_{\ell}}=-k_{\mu}k^{\mu}q_{\pm\hspace{0.5mm}\nu_{1}\cdots\nu_{\ell}}-\frac{2\epsilon{k}_{0}}{a^{2}}\sum_{j=1}^{n}g_{j}\int{d}\tau_{j}\hspace{0.5mm}\dot{u}_{j\nu_{1}}\cdots\dot{u}_{j\nu_{\ell}}\delta(x_{0}-u_{j0})\\
    \times\left[\mathrm{Re}(z)\sin(k_{\alpha}(x^{\alpha}-u^{\alpha}_{j}))-\mathrm{Im}(z)\cos(k_{\alpha}(x^{\alpha}-u^{\alpha}_{j}))\right]\label{EqReal2}
\end{multline}
by introducing the auxiliary variables
\begin{equation}
    q_{\pm\hspace{0.5mm}\nu_{1}\cdots\nu_{\ell}}=\pm{i}\epsilon\left([\mathrm{Re}(z)\pm{i}\mathrm{Im}(z)]\Tilde{T}_{\pm\hspace{0.5mm}\nu_{1}\cdots\nu_{\ell}}-[\mathrm{Re}(z)\mp{i}\mathrm{Im}(z)]\Tilde{T}_{\pm\hspace{0.5mm}\nu_{1}\cdots\nu_{\ell}}^{\ast}\right),\label{QTensor1}
\end{equation}
\begin{equation}
    \pi_{\pm\hspace{0.5mm}\mu\nu_{1}\cdots\nu_{\ell}}=\epsilon{k_{\mu}}\left([\mathrm{Re}(z)\pm{i}\mathrm{Im}(z)]\Tilde{T}_{\pm\hspace{0.5mm}\nu_{1}\cdots\nu_{\ell}}+[\mathrm{Re}(z)\mp{i}\mathrm{Im}(z)]\Tilde{T}_{\pm\hspace{0.5mm}\nu_{1}\cdots\nu_{\ell}}^{\ast}\right),\label{PiTensor1}
\end{equation}
with $z$ a scalar constant. The possibility of defining two distinct pairs of auxiliary variables is due to the fact that the only condition imposed on $a^2$ and $b^2$ is that their quotient is equal to $\kappa^2$. This multiplicity in the definitions of the auxiliary variables results helpful in operationally handling the oscillation modes with positive and negative frequencies in a symmetric manner. The factor $\epsilon$ has been introduced in the definitions for the sake of generality.

Consequently, based on the structures of Eqs. (\ref{EqReal1}) and (\ref{EqReal2}), which resemble the evolution equations of the canonical variables of a harmonic oscillator, we conclude that a function ${\mathcal{J}}$, such that
\begin{equation}
    \partial_{\mu}q_{\pm\hspace{0.5mm}\nu_{1}\cdots\nu_{\ell}}=\frac{\partial\mathcal{J}}{\partial\pi^{\mu\nu_{1}\cdots\nu_{\ell}}_{\pm}},\quad\partial^{\mu}\pi_{\pm\hspace{0.5mm}\mu\nu_{1}\cdots\nu_{\ell}}=-\frac{\partial\mathcal{J}}{\partial{q}^{\nu_{1}\cdots\nu_{\ell}}_{\pm}},\label{CanonicalDWEqsTensor2}
\end{equation}
exists and is given, up to a constant of integration and a global sign, by
\begin{multline}
    \mathcal{J}=\frac{1}{2}\left(\pi_{\pm\hspace{0.5mm}\mu\nu_{1}\cdots\nu_{\ell}}\pi^{\mu\nu_{1}\cdots\nu_{\ell}}_{\pm}+k_{\mu}k^{\mu}q_{\pm\hspace{0.5mm}\nu_{1}\cdots\nu_{\ell}}q^{\nu_{1}\cdots\nu_{\ell}}_{\pm}\right)+\frac{2\epsilon{k}_{0}}{a^{2}}\sum_{j=1}^{n}g_{j}\int{d}\tau_{j}\hspace{0.5mm}\dot{u}_{j\nu_{1}}\cdots\dot{u}_{j\nu_{\ell}}\delta(x_{0}-u_{j0})\\
    \times\left(\frac{k_{\mu}\pi_{\pm}^{\mu\nu_{1}\cdots\nu_{\ell}}}{k_{\alpha}k^{\alpha}}\left[\mathrm{Re}(z)\cos(k_{\alpha}(x^{\alpha}-u^{\alpha}_{j}))+\mathrm{Im}(z)\sin(k_{\alpha}(x^{\alpha}-u^{\alpha}_{j}))\right]\right.\\
    \left.\textcolor{white}{\bigg|}+q_{\pm}^{\nu_{1}\cdots\nu_{\ell}}\left[\mathrm{Re}(z)\sin(k_{\alpha}(x^{\alpha}-u^{\alpha}_{j}))-\mathrm{Im}(z)\cos(k_{\alpha}(x^{\alpha}-u^{\alpha}_{j}))\right]\right)\label{HamilEspMomCTensorQPi}
\end{multline}
or, equivalently, by
\begin{multline}
    \mathcal{J}=2\epsilon^{2}\vert{z}\vert^{2}{k}_{\mu}k^{\mu}\left(\Tilde{T}_{+\hspace{0.5mm}\nu_{1}\cdots\nu_{\ell}}^{\ast}\Tilde{T}_{+}^{\nu_{1}\cdots\nu_{\ell}}+\Tilde{T}_{-\hspace{0.5mm}\nu_{1}\cdots\nu_{\ell}}^{\ast}\Tilde{T}_{-}^{\nu_{1}\cdots\nu_{\ell}}\right)\\
    +\frac{2\epsilon^{2}\vert{z}\vert^{2}{k}_{0}}{a^{2}}\sum_{j=1}^{n}g_{j}\int{d}\tau_{j}\hspace{0.5mm}\dot{u}_{j\nu_{1}}\cdots\dot{u}_{j\nu_{\ell}}\delta(x_{0}-u_{j0})\left[\left(\Tilde{T}_{+}^{\nu_{1}\cdots\nu_{\ell}}+\Tilde{T}^{\ast\hspace{0.5mm}\nu_{1}\cdots\nu_{\ell}}_{-}\right)e^{ik_{\mu}(x^{\mu}-u^{\mu}_{j})}+\textrm{c.c.}\right].\label{HamilEspRecipTensor}
\end{multline}
Eqs. (\ref{HamilEspMomCTensorQPi}) and (\ref{HamilEspRecipTensor}) show that function $\mathcal{J}$ is a Lorentz-invariant, real function. We will show next that a connection between this function and the image of the Donder–Weyl function in Eq. (\ref{DWFunctionTensor}) in the momentum representation of the field can be established.

For this purpose, we now consider the plane-wave expansion of the free tensor field, that is, before it begins to interact with the particles, and proceed to obtain the image of the de Donder–Weyl function in the momentum representation by integrating Eq. (\ref{DWFunctionTensor}) over the whole space-time. Accordingly, we find that
\begin{equation}
    \int{d}^{4}x\hspace{0.5mm}\mathcal{H}=\frac{1}{8\pi^{3}}\int{d}^{4}k\hspace{0.5mm}\Theta(k_{0})\delta({k}_{\mu}k^{\mu}-\kappa^{2})\int{d}x_{0}\hspace{0.5mm}\Tilde{\mathcal{H}}_{\mathrm{free}},
\end{equation}
where the function $\Tilde{\mathcal{H}}_{\mathrm{free}}$ is
\begin{multline}
    \Tilde{\mathcal{H}}_{\hspace{0.25mm}\mathrm{free}}=\frac{b^{2}}{k_{0}}\left(\Tilde{T}_{+\hspace{0.5mm}\nu_{1}\cdots\nu_{\ell}}^{\hspace{0.25mm}\mathrm{free}\hspace{0.5mm}\ast}\Tilde{T}_{+\hspace{0.5mm}}^{\hspace{0.25mm}\mathrm{free}\hspace{0.5mm}\nu_{1}\cdots\nu_{\ell}}+\Tilde{T}_{-\hspace{0.5mm}\nu_{1}\cdots\nu_{\ell}}^{\hspace{0.25mm}\mathrm{free}\hspace{0.5mm}\ast}\Tilde{T}_{-\hspace{0.5mm}}^{\hspace{0.25mm}\mathrm{free}\hspace{0.5mm}\nu_{1}\cdots\nu_{\ell}}\right)\\
    +\sum_{j=1}^{n}g_{j}\int{d}\tau_{j}\hspace{0.5mm}\dot{u}_{j\nu_{1}}\cdots\dot{u}_{j\nu_{\ell}}\delta(x_{0}-u_{j0})\left[\left(\Tilde{T}_{+}^{\hspace{0.5mm}\mathrm{free}\hspace{0.5mm}\nu_{1}\cdots\nu_{\ell}}+\Tilde{T}^{\hspace{0.5mm}\mathrm{free}\hspace{0.5mm}\ast\hspace{0.5mm}\nu_{1}\cdots\nu_{\ell}}_{-}\right)e^{-ik_{\mu}u^{\mu}_{j}}+\textrm{c.c.}\right].\label{HamilEspRecipTensor2}
\end{multline}
Hence, by comparing Eq. (\ref{HamilEspRecipTensor}) with Eq. (\ref{HamilEspRecipTensor2}), we find that the two expressions coincide if 
\begin{equation}
    \epsilon^{2}=\frac{a^{2}}{2k_{0}\vert{z}\vert^{2}}\label{FigEpsilon}
\end{equation}
is assumed in Eq. (\ref{HamilEspRecipTensor}) and the substitutions
\begin{equation}
    \Tilde{T}^{\mathrm{free}}_{\pm\hspace{0.5mm}\nu_{1}\cdots\nu_{\ell}}\rightarrow\Tilde{T}_{\pm\hspace{0.5mm}\nu_{1}\cdots\nu_{\ell}}e^{\pm{i}k_{\mu}x^{\mu}},\quad\int{d}x_{0}\rightarrow\int{d}^{4}x\label{CorrespTensor}
\end{equation}
are implemented in Eq. (\ref{HamilEspRecipTensor2}). The assignment in Eq. (\ref{FigEpsilon}) fixes the scale of $\mathcal{J}$ to that of the system. In contrast to $\tilde{\mathcal{H}}_{\mathrm{free}}$, whose scale is determined by its direct derivation from the Donder–Weyl function, $\mathcal{J}$ does not possess a predetermined scale. Its value depends on the normalization chosen for $q_{\pm\hspace{0.5mm}\nu_{1}\cdots\nu_{\ell}}$ and $\pi_{\pm\hspace{0.5mm}\mu\nu_{1}\cdots\nu_{\ell}}$, and therefore includes an arbitrary multiplicative factor. The choice of $\epsilon^{2}$ in Eq. (\ref{FigEpsilon}) removes this freedom by fixing the normalization of the auxiliary variables consistently with the system. The substitutions in Eq. (\ref{CorrespTensor}), in turn, restore the dependence of $\tilde{\mathcal{H}}_{\mathrm{free}}$ on $x_{\mu}$. This dependence was lost in the derivation of the function because the calculation involved integration over the entire space-time and the use of the orthogonality properties of the exponential functions in the plane-wave expansion of the field. As a consequence, the image of the Donder–Weyl function in the momentum space depends only on $x_{0}$. These substitutions additionally allow extending the result in Eq. (\ref{HamilEspRecipTensor2}) from the free case to the interaction situation. Therefore, the prescriptions in Eqs. (\ref{FigEpsilon}) and (\ref{CorrespTensor}) make possible to express both functions, $\mathcal{J}$ and $\tilde{\mathcal{H}}_{\mathrm{free}}$, in forms that can be directly compared.


On this basis, we observe the following. First, the retarded solution of the field equation, typically obtained from the Euler–Lagrange equation and thus associated with the Lagrangian formalism, yields the evolution equations for the amplitudes $\Tilde{T}_{\pm\hspace{0.5mm}\nu_{1}\cdots\nu_{\ell}}$. These can be then decomposed into first-order differential equations for $q_{\pm\hspace{0.5mm}\nu_{1}\cdots\nu_{\ell}}$ and $\pi_{\pm\hspace{0.5mm}\mu\nu_{1}\cdots\nu_{\ell}}$. Once Eq. (\ref{FigEpsilon}) is taken into account, the resulting equations resemble covariant Hamilton's equations for the auxiliary variables and the Lorentz-invariant function $\mathcal{J}$, thereby reproducing the characteristic structures of the MHF. Second, after restoring the dependence on the space-time coordinates according to Eq. (\ref{CorrespTensor}) and introducing the definitions of $q_{\pm\hspace{0.5mm}\nu_{1}\cdots\nu_{\ell}}$ and $\pi_{\pm\hspace{0.5mm}\mu\nu_{1}\cdots\nu_{\ell}}$, the image of the Donder–Weyl function in the momentum representation of the field leads to covariant Hamilton's equations that imply, when recombined, the second-order differential equation
\begin{equation}
    (\partial_{\alpha}\partial^{\alpha}+\kappa^{2})\Tilde{T}_{\pm\hspace{0.5mm}\nu_{1}\cdots\nu_{\ell}}=-\frac{k_{0}}{a^{2}}\sum_{j=1}^{n}g_{j}\int{d}\tau_{j}\hspace{0.5mm}\dot{u}_{j\nu_{1}}\cdots\dot{u}_{j\nu_{\ell}}\delta(x_{0}-u_{j0})e^{\mp{i}k_{\mu}(x^{\mu}-u^{\mu}_{j})},
\end{equation}
which reproduces the field equation after being multiplied by $\Theta(k_0)\delta({k}_{\mu}k^{\mu}-\kappa^{2})/8\pi^{3}$ and integrated over the entire momentum space, that is, after fully restoring its space-time representation. Together, these observations allow us to conclude that in the momentum representation of the field, the Lagrangian and multisymplectic descriptions of the system are connected in accordance with
\begin{equation}
    \mathcal{H}=\frac{1}{8\pi^{3}}\int{d}^{4}k\hspace{0.5mm}\Theta(k_{0})\delta({k}_{\mu}k^{\mu}-\kappa^{2})\mathcal{J}.\label{ConnectionLtoMHF}
\end{equation}
Hence, function $\mathcal{J}$ is identified with the image of the Donder–Weyl function in momentum space, and Eqs. (\ref{CanonicalDWEqsTensor2}) correspond to the covariant Hamilton's equations for the variables $q_{\pm\hspace{0.5mm}\nu_{1}\cdots\nu_{\ell}}$ and $\pi_{\pm\hspace{0.5mm}\mu\nu_{1}\cdots\nu_{\ell}}$, which assume the role of canonical variables of the field.

\subsection{Example 1: The Complex Scalar Field}

One of the simplest systems for testing the description provided by the MHF is the complex scalar field. In that case, constants $a^{2}$ and $b^{2}$ become $s^{2}/c$ and $m^{2}c$, respectively, so that $\vert\kappa\vert={mc}/s$, with $s$ and $m$ proportionality constants having dimensions of action and mass, respectively. Moreover, Eq. (\ref{InterDensityTensor}) reduces to Bergmann's proposal for the interaction Lagrangian density \cite{OB1956}. Then, from Eq. (\ref{DefMomMulti}), we obtain that the multisymplectic momenta are
\begin{equation}
    \theta_{\mu}=\frac{s^{2}}{c}\partial_{\mu}\phi^{\ast},\label{DefMomScalar}
\end{equation}
and its complex conjugate, and, therefore, that the de Donder–Weyl function for the complex scalar field is given by
\begin{equation}
    \mathcal{H}=\frac{c}{s^{2}}\theta_{\mu}^{\ast}\theta^{\mu}+m^{2}c\vert\phi\vert^{2}+(\phi+\phi^{\ast})\sum_{j=1}^{n}g_{j}\int{d}\tau_{j}\hspace{0.5mm}\delta^{(4)}(x_{\alpha}-u_{j\alpha}),\label{HamilScalarComplex}
\end{equation}
which leads to inhomogeneous Klein–Gordon equations as field equations for $\phi$ and $\phi^{\ast}$ in full agreement with the Lagrangian formalism.

The procedure for solving the dynamical problem associated with the scalar field is analogous to that of the tensor case, so that it is straightforward to obtain that a function $\mathcal{J}$ exists for the complex scalar field, that this function is
\begin{equation}
    \mathcal{J}=\frac{m^{2}c}{k_{0}}\left(\vert\Tilde{\phi}_{+}\vert^{2}+\vert\Tilde{\phi}_{-}\vert^{2}\right)+\sum_{j=1}^{n}g_{j}\int{d}\tau_{j}\hspace{0.5mm}\delta(x_{0}-u_{j0})\left[\left(\Tilde{\phi}_{+}+\Tilde{\phi}^{\ast}_{-}\right)e^{ik_{\mu}(x^{\mu}-u^{\mu}_{j})}+\textrm{c.c.}\right],\label{HamilEspRecipScalarComplex}
\end{equation}
and that it can be identified with the image of the de Donder–Weyl function of Eq. (\ref{HamilScalarComplex}) in the momentum representation, since it corresponds to $\Tilde{\mathcal{H}}_{\mathrm{free}}$ when the substitutions in Eq. (\ref{CorrespTensor}) are taken into account for the scalar field.

\subsection{Example 2: The Electromagnetic Field}\label{CampEM}

Naturally, the most widely studied field theory is Maxwell's theory, which is characterized by the fact that the electromagnetic field is a real, massless vector field. Thus, we have that
\begin{equation}
    \mathcal{L}_{\mathrm{field}}=-\frac{1}{16\pi{c}}F_{\mu\nu}F^{\mu\nu}\label{LagranEMFreeDirect}
\end{equation}
and
\begin{equation}
    \mathcal{L}_{\mathrm{int}}=-\frac{1}{c^{2}}A_{\mu}J^{\mu},\label{LIntEMFieldDirect}
\end{equation}
with $F_{\mu\nu}=\partial_{\mu}A_{\nu}-\partial_{\nu}A_{\mu}$ the electromagnetic tensor, and $A_{\mu}$ and $J_{\mu}$ the usual potential and current density vectors, respectively. Unlike the cases considered above, the Maxwell Lagrangian density is singular, i.e.,
\begin{equation}
    \mathrm{det}\left(\frac{\partial^{2}\mathcal{L}}{\partial(\partial_{\mu}A_{\alpha})\partial(\partial_{\nu}A^{\alpha})}\right)=0.
\end{equation}
This singular character results from the well-known redundancy of $A_{\mu}$, which is encoded in the antisymmetric structure of $F_{\mu\nu}$ and manifested in the invariance of the Lagrangian density under local gauge transformations. As in the usual Hamiltonian formalism, a singular Lagrangian density implies that the Legendre transformation relating the field derivatives to the multisymplectic momenta is not invertible and that the system is therefore subject to constraints that prevent the field derivatives from being expressed entirely in terms of the corresponding momenta. 

If the methods of the MHF are nevertheless applied to the Maxwell Lagrangian density, the associated Legendre transformation appears to be invertible. The multisymplectic momentum conjugate for $A_{\mu}$ is
\begin{equation}
    \theta_{\mu\nu}=-\frac{1}{4\pi{c}}F_{\mu\nu},\label{DefMomEMDirect}
\end{equation}
and the de Donder–Weyl function is thus given by
\begin{equation}
    \mathcal{H}=-\pi{c}\theta_{\mu\nu}\theta^{\mu\nu}-\mathcal{L}_{\mathrm{int}}.\label{HamilEMDirect}
\end{equation}
However, one immediately finds that this function leads to the incorrect equation
\begin{equation}
    \partial_{\mu}A_{\nu}=-2\pi{c}\theta_{\mu\nu}.
\end{equation}
The reason of this failure is that the constraint $\theta_{\mu\nu}+\theta_{\nu\mu}=0$ holds, so only the antisymmetric part of the above equation is valid. In general, the symmetry properties of the fields or the gauge invariance of the Lagrangian density imply a degenerate Legendre transformation and hence lead to algebraic relations between field derivatives and multisymplectic momenta, or among components of the multisymplectic momenta themselves, as occurs in the electromagnetic case. These relations appear in the MHF as constraints that encode the redundancies of the theory \cite{Dirac2001,JC1991}. As will be discussed in Sec. \ref{Dirac}, such constrained systems can be treated consistently by introducing Lagrange multipliers.

Alternatively, we follow a different approach, standard in many formulations of quantum electrodynamics, by explicitly breaking the gauge invariance of the Maxwell Lagrangian density through the addition of a gauge-fixing term to $\mathcal{L}_{\mathrm{field}}$:
\begin{equation}
    \mathcal{L}_{\mathrm{field}}=-\frac{1}{16\pi{c}}(F_{\mu\nu}F^{\mu\nu}+2\zeta(\partial_{\mu}A^{\mu})^{2}),\label{LagranEMFree}
\end{equation}
where $\zeta$ is a positive, real parameter\footnote{The Lagrangian density in Eq. (\ref{LagranEMFree}) differs from that obtained by specializing Eq. (\ref{LTensorField}) to the electromagnetic field, the latter corresponding to the so-called Fermi Lagrangian and being associated with the Lorentz gauge. For $\zeta=1$, the two densities differ only by a term whose space-time integral vanishes upon imposing the Lorentz condition. Therefore, Eq. (\ref{LagranEMFree}) constitutes a more general Lagrangian density for the free electromagnetic field.}. This term removes the degeneracy of the Legendre transformation associated with the gauge invariance and implies the multisymplectic momentum
\begin{equation}
    \theta_{\mu\nu}=-\frac{1}{4\pi{c}}(F_{\mu\nu}+\zeta\eta_{\mu\nu}\partial_{\alpha}A^{\alpha})\label{DefMomEM}
\end{equation}
and the de Donder–Weyl function
\begin{equation}
    \mathcal{H}=-\frac{1}{2}\pi{c}\theta_{\mu\nu}\left[\left(1+\frac{1}{2\zeta}\right)\theta^{\mu\nu}-\left(1-\frac{1}{2\zeta}\right)\theta^{\nu\mu}\right]-\mathcal{L}_{\mathrm{int}},\label{HamilEM}
\end{equation}
which yields two covariant Hamilton's equations that, in turn, lead to a field equation that is equivalent to well-known Maxwell equations for the electromagnetic vector potential in the Lorentz gauge.

Then, from the method of Green’s functions, we have that the retarded solution of the field equation is
\begin{equation}
    A_{\mu}=\frac{1}{8\pi^{3}}\lim_{m\rightarrow0}\left[\delta^{\nu}_{\mu}+\frac{\zeta-1}{\zeta}\frac{d}{dm^{2}}\partial_{\mu}\partial^{\nu}\right]\int{d}^{4}k\hspace{0.5mm}\Theta(k_{0})\delta(k_{\alpha}k^{\alpha}-m^{2})\tilde{A}_{\nu}+\mathrm{c.c.},
\end{equation}
where $\tilde{A}_{\mu}$ is a vector amplitude defined by
\begin{equation}
    \tilde{A}_{\mu}=\left(\tilde{A}_{\mu}^{\mathrm{free}}+4\pi{i}\sum_{j=1}^{n}e_{j}\int{d}\tau_{j}\hspace{0.5mm}\dot{u}_{j\mu}\Theta(x_{0}-u_{j0})e^{ik_{\alpha}u^{\alpha}_{j}}\right)e^{-ik_{\alpha}x^{\alpha}},
\end{equation}
with $e_{j}$ the electric charge of the $j$-th point particle. Now, by recalling that the purpose of the present work is to set the groundwork for generalizing Cetto's approach to quantization to the classical electromagnetic field, we shall limit ourselves from this point on to working in the Lorentz gauge, i.e., taking $\zeta=1$. Hence, we have that $\Tilde{A}_{\nu}$ satisfies the first-order differential equation
\begin{equation}
    (k_{\alpha}\partial^{\alpha}+i{k}_{\alpha}k^{\alpha})\Tilde{A}_{\nu}=4\pi{i}k_{0}\sum_{j=1}^{n}e_{j}\int{d}\tau_{j}\hspace{0.5mm}\dot{u}_{j\nu}\delta(x_{0}-u_{j0})e^{-ik_{\mu}(x^{\mu}-u^{\mu}_{j})}.\label{EcAmplVectorEM}
\end{equation}
So, by defining the auxiliary variables $q_{\mu}$ and $\pi_{\mu\nu}$ as\footnote{An extra multiplicative factor $2^{1/2}$ has been introduced in the definitions of $q_{\mu}$ and $\pi_{\mu\nu}$ with respect to Eqs. (\ref{PiTensor1}) and (\ref{QTensor1}), since the appropriate value for $\epsilon$ would include such a scaling correction, if those equations were used directly for real fields.}
\begin{equation}
    q_{\mu}=\frac{i(z\Tilde{A}_{\mu}-z^{\ast}\Tilde{A}_{\mu}^{\ast})}{\vert{z}\vert\sqrt{8\pi{c}{k}_{0}}},\quad\pi_{\mu\nu}=\frac{k_{\mu}(z\Tilde{A}_{\nu}+z^{\ast}\Tilde{A}_{\nu}^{\ast})}{\vert{z}\vert\sqrt{8\pi{c}{k}_{0}}},
\end{equation}
respectively, we decompose Eq. (\ref{EcAmplVectorEM}) into a pair of first-order differential equations with real coefficients, and, by assuming that there is a function $\mathcal{J}$ that implies Eqs. (\ref{CanonicalDWEqsTensor2}), we obtain that such function is given by
\begin{equation}
    \mathcal{J}=-\frac{{k}_{\mu}k^{\mu}}{4\pi{c}{k}_{0}}\Tilde{A}^{\ast}_{\nu}\Tilde{A}^{\nu}+\frac{1}{c}\sum_{j=1}^{n}e_{j}\int{d}\tau_{j}\hspace{0.5mm}\dot{u}_{j\nu}\delta(x_{0}-u_{j0})\left[\Tilde{A}^{\nu}e^{ik_{\mu}(x^{\mu}-u^{\mu}_{j})}+\textrm{c.c.}\right],\label{HamilEspRecipEM}
\end{equation}
and that it can be identified with the image of the de Donder–Weyl function for the electromagnetic field in the momentum representation.

The fact that the electromagnetic field is non-massive has as an immediate consequence that the field's dispersion relation is reduced to ${k}_{\mu}k^{\mu}=0$ and, therefore, that the first term in the right-hand side of Eq. (\ref{HamilEspRecipEM}) is identically zero. The null value of the free part of the de Donder–Weyl function will not be of major concern, since what matters about any Hamiltonian density is its functional dependence on the canonical variables, not its absolute value.

\section{Scope of the MHF: the Classical Dirac Field}\label{Dirac}

A system composed of spinor fields constitutes an ideal opportunity to explore the scope of the MHF, since spinors have different transformation rules under Lorentz transformations than those associated with usual tensor representations of the Lorentz group. Hence, let us now consider the classical Dirac field, which is described by the free propagation Lagrangian density
\begin{equation}
    \mathcal{L}_{\mathrm{field}}=\frac{is}{2}(\Bar{\psi}\gamma^{\mu}\partial_{\mu}\psi-\partial_{\mu}\Bar{\psi}\gamma^{\mu}{\psi})-mc\Bar{\psi}\psi,\label{LDiracField}
\end{equation}
where $\gamma^{\mu}$ denotes Dirac matrices, $\Bar{\psi}$ represents the Dirac-conjugate of $\psi$, and $s$ and $m$ are once again proportionality constants with dimensions of action and mass, respectively.

To establish a coupling between point particles and the classical Dirac field that gives rise in the field equation to an inhomogeneity with the same structure as that analyzed in Sect. \ref{DevelopGral}, we recall that quadratic forms $\Bar{\psi}_{1}{\psi}_{2}$, $\Bar{\psi}_{1}\gamma^{\mu}{\psi}_{2}$, and $\Bar{\psi}_{1}\gamma^{\mu}\gamma^{\nu}{\psi}_{2}$ transform under Lorentz transformations in accordance with tensor representations of the Lorentz group for ${\psi}_{1}$ and ${\psi}_{2}$ arbitrary spinors, so we propose as interaction Lagrangian density
\begin{equation}
    \mathcal{L}_{\mathrm{int}}=-\left(\sum_{j=1}^{n}\int{d}\tau_{j}\hspace{0.5mm}\Bar{\xi}_{j}\delta^{(4)}(x_{\alpha}-u_{j\alpha})\right)\psi-\mathrm{h.c.},\label{IntDiracSca}
\end{equation}
where
\begin{equation}
    {\xi}_{j}={\xi}_{1j}+\dot{u}_{j\mu}\gamma^{\mu}{\xi}_{2j}+\dot{u}_{j\mu}\dot{u}_{j\nu}\gamma^{\mu}\gamma^{\nu}{\xi}_{3j},
\end{equation}
with spinors $\xi_{ij}$, $i=1,2,3$, having the role of coupling constants of the $j$-th point particle with the field. As a result, point particles interact with the classical Dirac field through three different mechanisms\footnote{The Lagrangian density in Eq. (\ref{IntDiracSca}) is not related to the coupling term used in quantum electrodynamics for describing the interaction between photons and 1/2-spin fermions.}.

The direct application of Eq. (\ref{MultiMomentumDef}) results in two multisymplectic momenta for the classical Dirac field that are given by
\begin{equation}
    \theta_{\mu}=-\frac{is}{2}\gamma_{\mu}\psi\label{MultMomDirac1}
\end{equation}
and its Dirac-conjugate. These momenta have a completely different nature than those previously obtained, since the former ones are proportional to the fields themselves. Thus, Eq. (\ref{MultMomDirac1}), rather than a definition, represents a primary constraint in phase space that relates fields to multisymplectic momenta:
\begin{equation}
    \phi_{\mu}(\psi,\theta_{\alpha})=\theta_{\mu}+\frac{is}{2}\gamma_{\mu}\psi=0.
\end{equation}
As stated above, this constraint arises from the singular character of the Lagrangian density of the classical Dirac field, as indicated by the vanishing determinant of its Hessian matrix,
\begin{equation}
    \mathrm{det}\left(\frac{\partial^{2}\mathcal{L}}{\partial(\partial_{\mu}\psi)\partial(\partial_{\nu}\psi)}\right)=0.
\end{equation}
Nonetheless, by exploiting the formal analogy between the Hamiltonian formulation of classical mechanics and the MHF, we can follow Dirac’s prescription for handling singular Lagrangians and introduce Lagrange multipliers \cite{Dirac2001}. Accordingly, we have that the de Donder–Weyl function is given by\footnote{Similar to the procedure followed in Sec. \ref{CampEM}, Struckmeier and Redelbach address the singular character of the Lagrangian density of the classical Dirac field at the level of the dynamics by adding the divergence-free term $\partial_{\mu}\Bar{\psi}\hspace{0.25mm}[\gamma^{\mu},\gamma^{\nu}]\hspace{0.25mm}\partial_{\nu}\psi$. This modification leaves the equations of motion unchanged while rendering the Legendre transformation nondegenerate \cite{JS2008}.}
\begin{equation}
    \mathcal{H}'=\mathcal{H}+\Bar{\chi}_{\mu}\phi^{\mu}(\psi,\theta_{\alpha})+\Bar{\phi}_{\mu}(\Bar{\psi},\Bar{\theta}_{\alpha}){\chi}^{\mu},
\end{equation}
where $\chi_{\mu}$ is an undetermined spinor multiplier and $\mathcal{H}$ corresponds to the de Donder–Weyl function introduced in Eq. (\ref{LegendreTrans2}). This automatically leads to
\begin{equation}
    \mathcal{H}'=\Bar{\chi}_{\mu}\left(\theta^{\mu}+\frac{is}{2}\gamma^{\mu}\psi\right)+\left(\Bar{\theta}^{\mu}-\frac{is}{2}\Bar{\psi}\gamma^{\mu}\right)\chi_{\mu}+mc\Bar{\psi}\psi-\mathcal{L}_{\mathrm{int}}.\label{DeDonWeylSpinor}
\end{equation}
We emphasize that the de Donder–Weyl functions $\mathcal{H}$ and $\mathcal{H}'$ are identical because the constraint $\phi_{\mu}(\psi,\theta_{\alpha})=0$ holds. However, $\mathcal{H}'$ admits independent variations of $\psi$ and $\theta_{\mu}$, whereas $\mathcal{H}$ does not \cite{Dirac2001}. Thereby, the covariant Hamilton's equations in Eqs. (\ref{EqHamiltonPi}) imply equations
\begin{equation}
    \chi_{\mu}=\partial_{\mu}{\psi},\label{IndMultDirac}
\end{equation}
which determines the Lagrange multipliers,
\begin{equation}
    \partial_{\mu}\theta^{\mu}=-mc{\psi}+\frac{is}{2}\gamma^{\mu}\chi_{\mu}-\sum_{j=1}^{n}\int{d}\tau_{j}\hspace{0.5mm}{\xi}_{j}\delta^{(4)}(x_{\alpha}-u_{j\alpha}),\label{EcDirac1}
\end{equation}
and their Dirac-conjugates. So, by substituting Eqs. (\ref{MultMomDirac1}) and (\ref{IndMultDirac}) into Eq. (\ref{EcDirac1}), we obtain that the field equation for $\psi$ is, indeed, an inhomogeneous Dirac equation.

Once again, we obtain from the method of Green’s functions that the complex spinor amplitudes in the retarded solution of the field equation satisfy the first-order differential equations
\begin{equation}
    (k_{\alpha}\partial^{\alpha}\pm{i}{k}_{\alpha}k^{\alpha})\Tilde{\psi}_{\pm}=\mp\frac{ik_{0}}{s}(\kappa\pm{k}_{\alpha}\gamma^{\alpha})\sum_{j=1}^{n}\int{d}\tau_{j}\hspace{0.5mm}{\xi}_{j}\Theta(x_{0}-u_{j0})e^{\mp{i}k_{\mu}(x^{\mu}-u^{\mu}_{j})}.
\end{equation}
Hence, by applying the definitions of the real auxiliary variables in Eqs. (\ref{PiTensor1}) and (\ref{QTensor1}) to the spinor case and assuming that there exists a function $\mathcal{J}$ which satisfies Eqs. (\ref{CanonicalDWEqsTensor2}), we find after a long, but straightforward calculation that such function exists and can be expressed, up to a constant of integration and a global sign, as
\begin{equation}
    \mathcal{J}=\mathcal{J}_{1}(\tilde{\psi}_{+},\tilde{\psi}_{-})+\mathcal{J}_{2}(\tilde{\psi}_{+}^{\ast},\tilde{\psi}_{-}^{\ast}),\label{JDirac}
\end{equation}
with $\mathcal{J}_{1}$ defined by
\begin{multline}
    \mathcal{J}_{1}(\tilde{\psi}_{+},\tilde{\psi}_{-})=2\epsilon^{2}\vert{z}\vert^{2}\kappa^{2}\left(\bar{\tilde{\psi}}_{+}\tilde{\psi}_{+}+\bar{\tilde{\psi}}_{-}\tilde{\psi}_{-}\right)+\frac{2\epsilon^{2}\vert{z}\vert^{2}{k}_{0}}{s}\sum_{j=1}^{n}\int{d}\tau_{j}\hspace{0.5mm}\Theta(x_{0}-u_{j0})\\
    \times\left(\bar{\xi}_{j}\left[(\kappa+{k}_{\nu}\gamma^{\nu})\tilde{\psi}_{+}e^{ik_{\mu}(x^{\mu}-u^{\mu}_{j})}+(\kappa-{k}_{\nu}\gamma^{\nu})\tilde{\psi}_{-}e^{-ik_{\mu}(x^{\mu}-u^{\mu}_{j})}\right]+\mathrm{h.c.}\right),
\end{multline}
and $\mathcal{J}_{2}$ defined in a similar fashion.

In this scenario, we are again able to identify function $\mathcal{J}_{1}$ with the image of the de Donder–Weyl function in Eq. (\ref{DeDonWeylSpinor}) in the momentum representation if we set $\epsilon^{2}=s/4k_{0}\kappa\vert{z}\vert^{2}$ and take into account the substitutions in Eq. (\ref{CorrespTensor}). The difference in the values obtained for $\epsilon$ between Sects. \ref{DescripMomSpace} and \ref{Dirac} illustrates not only the fact that the field theory of the classical Dirac field emerges from the “linearization” of the Klein–Gordon equation, so that $a^{2}=s$, $b^{2}=mc$, and $b^{2}/a^{2}=\kappa$; but also that the complex amplitudes $\tilde{\psi}_{\pm}$ include an additional matrix factor associated with the idempotent matrix $(\kappa\pm{k}_{\mu}\gamma^{\mu})/2\kappa$. 

Another noteworthy feature in the analysis of the classical Dirac field is that Eq. (\ref{JDirac}) possesses two parts: one associated with $\tilde{\psi}_{\pm}$, $\mathcal{J}_{1}$, and another associated with $\tilde{\psi}^{\ast}_{\pm}$, $\mathcal{J}_{2}$; while the plane-wave expansion of the free field leads to a function that can be only identified with $\mathcal{J}_{1}$. This is only an apparent discrepancy because complex conjugate amplitudes $\tilde{\psi}_{\pm}^{\ast}$ were solely introduced for defining $\pi_{\pm\hspace{0.5mm}\mu}$ and $q_{\pm}$ as real spinors, not because they appeared in the field's Lagrangian density. Consequently, function $\mathcal{J}_2$ in Eq. (\ref{JDirac}) does not represent any additional dynamics; it only leads to the complex conjugate equations of motion for amplitudes $\tilde{\psi}_{\pm}$.

\section{Concluding Remarks}

\noindent The preceding results exhibit the descriptive power of the MHF a step beyond its basic theoretical framework by demonstrating the possibility of transitioning its description to the momentum representation and, additionally, support the interpretation of Eqs. (\ref{CanonicalDWEqsTensor2}) as the covariant Hamilton's equations for the canonical variables of the fields. Altogether, these results indicate that the incompatibility between the requirement of manifest covariance and the canonical formalism may, in principle, be resolved within the MHF.

Now, the next step toward extending Cetto’s approach to quantization and, therefore, toward formulating a Lorentz-covariant, canonical quantization scheme is the introduction of a Poisson bracket within the MHF. Nevertheless, as stated by Forger and his collaborators \cite{MF2002,MF2005,MF2015}, the status of such a construction remains unsatisfactory, since the formalism admits several alternative definitions of Poisson brackets. Marsden et al. proposed, for instance, that the bracket
\begin{equation}
    \{F,G\}_{\mathrm{Marsden}}=\int{d}^{4}x\hspace{0.5mm}V_{\mu}\left[\frac{\partial{F}}{\partial{A}^{\nu}}\frac{\partial{G}}{\partial\theta_{\mu\nu}}-\frac{\partial{G}}{\partial\theta_{\mu\nu}}\frac{\partial{F}}{\partial{A}^{\nu}}\right]\label{ParentesisMarsden}
\end{equation}
could serve as the Poisson bracket of variables 
$F$ and $G$ with respect to the electromagnetic vector potential and the multisymplectic momentum, where $V_{\mu}$ denotes an arbitrary space-time vector that is related to the passage to the equations of motion \cite{JM1986}, although its precise nature was not further specified.
On the other hand, von Hippel and Wohlfarth suggested defining the same object as
\begin{equation}
    \{F,G\}_{\mathrm{von}\hspace{0.5mm}\mathrm{Hippel}}=V_{\mu}\left[\frac{\partial{F}}{\partial{A}^{\nu}}\frac{\partial{G}}{\partial\theta_{\mu\nu}}-\frac{\partial{G}}{\partial\theta_{\mu\nu}}\frac{\partial{F}}{\partial{A}^{\nu}}\right],
\end{equation}
where $V_{\mu}$ is taken to be a vector proportional to Dirac matrices. This construction, however, reproduces the results of standard canonical quantization only for one-dimensional parameter spaces \cite{VG2006}. Another noteworthy proposal was put forward by Struckmeier and Redelbach \cite{JS2008}, who introduced the bracket
\begin{equation}
    (\{F,G\}_{\mathrm{Struckmeier}})_{\mu}=\frac{\partial{F}}{\partial{A}_{\nu}}\frac{\partial{G}}{\partial\theta^{\mu\nu}}-\frac{\partial{G}}{\partial\theta^{\mu\nu}}\frac{\partial{F}}{\partial{A}_{\nu}},
\end{equation}
which carries a free index and fails to satisfy the standard Jacobi identity when terms associated with different indices are considered.

We will alternatively identify the Poisson bracket with the bilinear form that naturally emerges from the formalism when analyzing the derivatives of the de Donder–Weyl function. Accordingly, one finds in the electromagnetic case that
\begin{equation}
    \partial_{\lambda}\mathcal{H}=\partial_{\lambda}\mathcal{H}\hspace{0.5mm}\Big|_{\mathrm{explicit}}+\frac{1}{8\pi^{3}}\int{d}^{4}k\hspace{0.5mm}\Theta(k_{0})\delta({k}_{\alpha}k^{\alpha})\left[\frac{\partial\mathcal{J}}{\partial{q}_{\nu}}\partial_{\lambda}q_{\nu}+\frac{\partial\mathcal{J}}{\partial\pi_{\mu\nu}}\partial_{\lambda}\pi_{\mu\nu}\right].\label{EstrucAlg}
\end{equation}
By recalling the covariant Hamilton's equations of $q_{\mu}$ and $\pi_{\mu\nu}$, this last equation can be then rewritten as
\begin{equation}
    \partial_{\lambda}\mathcal{H}=\partial_{\lambda}\mathcal{H}\hspace{0.5mm}\Big|_{\mathrm{explicit}}+\frac{1}{8\pi^{3}}\int{d}^{4}k\hspace{0.5mm}\Theta(k_{0})\delta({k}_{\alpha}k^{\alpha})\left[\partial_{\lambda}\pi^{\mu\nu}-\delta^{\mu}_{\lambda}\partial_{\gamma}\pi^{\gamma\nu}\right]\partial_{\mu}q_{\nu}.\label{EstrucAlg2}
\end{equation}
This result indicates that the de Donder–Weyl function does not, in general, constitute a conserved quantity. It is worth noting, however, that by employing the standard definition of the energy-momentum tensor and Eqs. (\ref{MultiMomentumDef}) and (\ref{LegendreTrans2}), one can relate the divergence of this tensor to the explicit derivative of the de Donder–Weyl function. This relation generalizes the identity between the total and explicit time derivatives of constants of motion from classical mechanics and indicates that when the de Donder–Weyl function lacks explicit dependence on space-time coordinates, the columns of the energy-momentum tensor constitute conserved currents. This observation was extended by Struckmeier and Redelbach, who derived a generalized form of Noether’s theorem and showed the fundamental role of the de Donder–Weyl function in the identification of conserved quantities associated with symmetries of the system \cite{JS2008}.

Although the de Donder–Weyl function is not, strictly speaking, a conserved quantity, Eq. (\ref{EstrucAlg}) does suggest identifying the bilinear, antisymmetric operator
\begin{equation}
    \lbrace{F},G\rbrace=\int{d}^{4}k\hspace{0.5mm}V_{\mu}\left[\frac{\partial{F}}{\partial{q}^{\nu}}\frac{\partial{G}}{\partial\pi_{\mu\nu}}-\frac{\partial{G}}{\partial\pi_{\mu\nu}}\frac{\partial{F}}{\partial{q}^{\nu}}\right],\label{Poisson}
\end{equation}
where $V_{\mu}$ denotes a vector in momentum space that has not yet been specified, with the Poisson bracket of variables $F$ and $G$ with respect to the canonical variables of the electromagnetic field. The introduction of $V_{\mu}$ is crucial, since it ensures both the Lorentz-invariant character of the proposal and the fulfillment of the Jacobi identity. Certainly, the proposed bracket possesses the same basic algebraic structure as those previously mentioned, but differs in that it is constructed in terms of the canonical variables of the fields rather than the fields themselves and their corresponding multisymplectic momenta. A more detailed analysis, based on the relation established in Eq. (\ref{ConnectionLtoMHF}), suggests that one might expand $A_{\mu}$ and $\theta_{\mu\nu}$ into normal modes of oscillation and employ this decomposition to recast Eq. (\ref{ParentesisMarsden}) in the form of Eq. (\ref{Poisson}). Doing so, however, would require assigning a definite expression to $V_{\mu}$, whose explicit form remains unknown at this stage.

The proposal in Eq. (\ref{Poisson}) subsequently calls for the promotion of vector $\pi_{\nu}=\pi_{\mu\nu}V^{\mu}$ as the conjugate momentum associated with $q_{\mu}$, since their Poisson bracket,
\begin{equation}
    \lbrace{q}_{\mu}(k_{\alpha}),\pi_{\nu}(k_{\alpha}')\rbrace=V_{\lambda}V^{\lambda}\eta_{\mu\nu}\delta^{(4)}(k_{\alpha}-k_{\alpha}'),\label{PoissonQPi}
\end{equation}
resembles the canonical Poisson bracket of classical (non-relativistic) mechanics if $V_{\mu}$ is chosen appropriately. This resemblance suggests a possible strategy for determining $V_{\mu}$: one may derive the non-covariant form of Eq. (\ref{PoissonQPi}) and compare it with the corresponding bracket in classical mechanics. Moreover, other brackets could be evaluated, for example, that relating the field at two different positions, and contrasted with the results obtained from the standard formulations of field theories, since those brackets must not depend on the auxiliary variables used in the description. More significantly, Eq. (\ref{PoissonQPi}) hints at the possibility of introducing a single Poisson bracket defined in terms of the canonical variables of the complete system, namely fields$+$particles. Such a Lorentz-covariant object would, in principle, permit a unified, Lorentz-covariant, canonical treatment of the system and could provide a solid foundation for developing a Lorentz-covariant, canonical quantization scheme. A more exhaustive discussion of the proposed description will be presented elsewhere.
\vspace{0.5cm}

\noindent \textbf{Acknowledgments}
\vspace{0.2cm}

\noindent The author gratefully acknowledges Professors A. M. Cetto and L. de la Peña for their continuous guidance throughout the development of this work and for their many helpful suggestions. The author also thanks three anonymous reviewers for their valuable comments that contributed to improving the manuscript.

\end{document}